\documentclass[prl,twocolumn,aps,superscriptaddress,reprint]{revtex4-1}

\usepackage{amsfonts}
\usepackage{mathrsfs}
\usepackage{amsmath}% needed for subequations
\usepackage{color}
\usepackage{graphicx}
\usepackage{bm}% bold maths
\usepackage{amssymb}
\usepackage{xspace}
\usepackage{epstopdf}
\usepackage{dcolumn}% Align table columns on decimal point
\usepackage{longtable}
\usepackage{multirow}
\usepackage[colorlinks=true, letterpaper=true, pdfstartview=FitV, linkcolor=blue, citecolor=blue, urlcolor=blue]{hyperref}

\begin{document}
\title{Hybrid Dirac Semimetal in CaAgBi Materials Family}

\author {Cong Chen}
\affiliation{Department of Physics, Key Laboratory of Micro-nano Measurement-Manipulation and Physics (Ministry of Education), Beihang University, Beijing 100191, China}
\affiliation{Research Laboratory for Quantum Materials, Singapore University of Technology and Design, Singapore 487372, Singapore}

\author{Shan-Shan Wang}
\affiliation{Research Laboratory for Quantum Materials, Singapore University of Technology and Design, Singapore 487372, Singapore}

\author{Lei Liu}
\affiliation{State Key Laboratory of Integrated Service Networks, Xidian University, Xi'an, China}

\author{Zhi-Ming Yu}\email{zhiming\_yu@sutd.edu.sg}
\affiliation{Research Laboratory for Quantum Materials, Singapore University of Technology and Design, Singapore 487372, Singapore}

\author {Xian-Lei Sheng}\email{xlsheng@buaa.edu.cn}
\affiliation{Department of Physics, Key Laboratory of Micro-nano Measurement-Manipulation and Physics (Ministry of Education), Beihang University, Beijing 100191, China}
\affiliation{Research Laboratory for Quantum Materials, Singapore University of Technology and Design, Singapore 487372, Singapore}

\author {Ziyu Chen}
\affiliation{Department of Physics, Key Laboratory of Micro-nano Measurement-Manipulation and Physics (Ministry of Education), Beihang University, Beijing 100191, China}

\author{Shengyuan A. Yang}
%\email{shengyuan\_yang@sutd.edu.sg}
\affiliation{Research Laboratory for Quantum Materials, Singapore University of Technology and Design, Singapore 487372, Singapore}

\begin{abstract}
Based on their formation mechanisms, Dirac points in three-dimensional systems can be classified as accidental or essential. The former can be further distinguished into type-I and type-II, depending on whether the Dirac cone spectrum is completely tipped over along certain direction. Here, we predict the coexistence of all three kinds of Dirac points in the low-energy band structure of CaAgBi-family materials with a stuffed Wurtzite structure. Two pairs of accidental Dirac points reside on the rotational axis, with one pair being type-I and the other pair type-II; while another essential Dirac point is pinned at the high symmetry point on the Brillouin zone boundary. Due to broken inversion symmetry, the band degeneracy around accidental Dirac points is completely lifted except along the rotational axis, which may enable the splitting of chiral carriers at a ballistic p-n junction with a double negative refraction effect. We clarify their symmetry protections, and find both the Dirac-cone and Fermi arc topological surface states.
\end{abstract}

%\pacs{71.20.-b, 73.20.-r, 31.15.A-}
\maketitle

%%% pacs
%%% 71.20.-b	Electron density of states and band structure of crystalline solids
%%% 73.20.-r	Electron states at surfaces and interfaces
%%% 31.15.A-	Ab initio calculations

Topological semimetals have been attracting tremendous interest in current research~\cite{RevModPhys.88.035005,Dai:2016bu,Zhao2013c}, partly because they offer a convenient platform to explore the intriguing physics of high-energy elementary particles. For example, Weyl semimetals possess linearly-dispersing two-fold-degenerate Weyl points close to Fermi energy ~\cite{WanXG_Weyl,1367-2630-9-9-356,Balents_2011Weyl,WengHM_2015TaAs,Huang:2015ic,Lv2015,Xu613}. Each Weyl point has a definite chirality of $\pm 1$, around which the quasiparticle excitations mimic the relativistic Weyl fermions~\cite{Volovik2003,NIELSEN1983389}.
Two Weyl points with opposite chirality would be unstable towards gap-opening when they meet at the same $k$-point, unless there exists additional symmetry protection, such as the case in so-called Dirac semimetals~\cite{PhysRevLett.108.140405}, in which stable Dirac points with four-fold-degeneracy make it possible to simulate massless Dirac fermions.

\begin{figure}[b!]
\includegraphics[width=8.6cm]{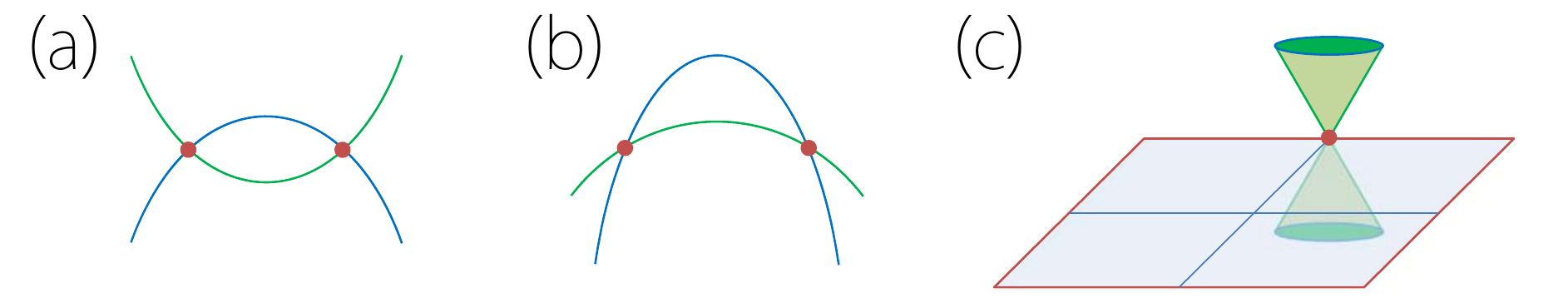}
\caption{Three kinds of Dirac points. (a) Type-I and (b) type-II accidental Dirac points. Here each band is two-fold degenerate. (c) Essential Dirac point located at high-symmetry point on the Brillouin zone boundary.}
\label{fig1}
\end{figure}

Dirac points can be classified as accidental or essential~\cite{Yang:2014ia} (see Fig.~\ref{fig1}). Accidental Dirac points require band inversion (which is in a sense accidental) and are stablized by certain symmorphic crystalline symmetries such as rotation, so such Dirac points typically reside on high-symmetry lines. Examples include the experimentally confirmed Dirac semimetals Na$_3$Bi~\cite{PhysRevB.85.195320,Liu864} and Cd$_3$As$_2$~\cite{PhysRevB.88.125427,Liu2014Exp_Cd3As2,Neupane:2014kc,PhysRevLett.113.027603}. Essential Dirac points do not need band inversion, and their presence is solely determined by certain nonsymmorphic symmtries at high-symmetry points on the boundary of the Brillouzin zone (BZ). Examples include the first few proposals such as $\beta$-cristobalite BiO$_2$ ~\cite{PhysRevLett.108.140405} and several Bi-containing distorted spinels~\cite{PhysRevLett.112.036403}. Accidental Dirac points can be removed by reverting the band ordering without changing the symmetry, whereas essential Dirac points cannot. In addition, since accidental Dirac points are located at $k$-points with lower symmetry (compared with essential ones), the dispersions around them are less constrained. It is possible to have the Dirac cone completely tipped over along certain direction~\cite{Soluyanov:2015cn,PhysRevLett.115.265304}, realizing a so-called type-II Dirac point~\cite{2016arXiv160607555C}, which has been identified in VAl$_3$~\cite{2016arXiv160607555C}, PdTe$_2$~\cite{2016arXiv161108112F,2017arXiv170505708X,2016arXiv161206946N}, PtSe$_2$ family~\cite{2017arXiv170304242Z,PhysRevB.94.121117,2016arXiv160703643Y} and others~\cite{PhysRevB.95.155112,2016arXiv160605042L,2017arXiv170309899W}. Fascinating yet distinct physics have been proposed for type-I and type-II points, and it has been theoretically argued in the context of Weyl points that when both types coexist in a single hybrid material, even more interesting effects could appear~\cite{PhysRevB.94.121105,PhysRevA.95.033629}.

In this work, we show that all the above-mentioned three kinds of Dirac points are simultaneously present in the low-energy band structure of CaAgBi-family materials. Two pairs of accidental Dirac points are realized on the $k_z$-axis and are protected by the $C_{6v}$ point group symmetry, with one pair being type-I and the other pair type-II. Meanwhile, another single essential Dirac point occurs at the A point on the boundary of BZ, as dictated by the nonsymmorphic space group symmetry. The inversion symmetry, preserved in most Dirac semimetals proposed so far, is broken in the CaAgBi-family materials. Consequently, the dispersions around the Dirac points exhibit distinct features. While each band is two-fold degenerate around the essential Dirac point, the band degeneracy around the accidental Dirac points is lifted due to the broken inversion symmetry (except along the rotational axis), making it possible to split the chiral carriers at a ballistic p-n junction with a double negative refraction effect. We clarify the symmetry protection mechanisms, and find interesting surface-orientation-dependent topological surface states including both the Dirac-cone surface bands and the surface Fermi arcs.  This class of hybrid Dirac materials offers a promising platform to explore the intriguing physics of Dirac fermions.

\begin{figure}[t!]
\includegraphics[width=9cm]{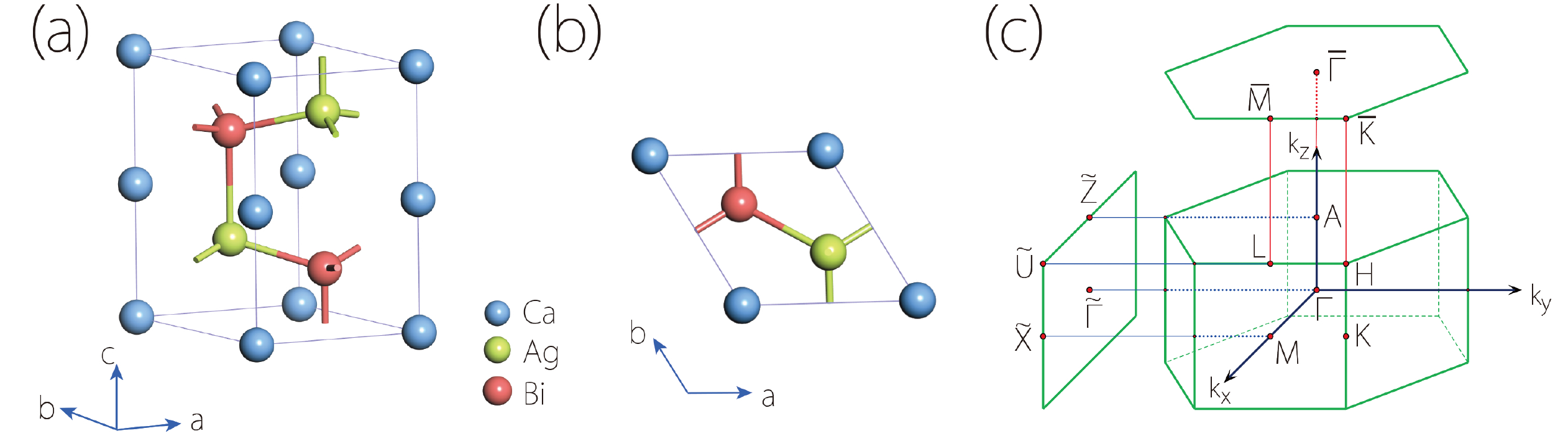}
\caption{(a) Side view and (b) top view of the crystal structure of CaAgBi. (c) Corresponding Brillouin zone. }
\label{fig2}
\end{figure}

The CaAgBi-family materials take a stuffed Wurtzite-type structure, with space group No. 186 ($P6_3mc$)~\cite{Exp_CaAgBi}, as shown in Fig.~\ref{fig2}. The structure may be viewed as a zinc-blende structure compressed along the [111]-direction. Since members of this family show similar low-energy band features, we shall focus on CaAgBi as a representative in the following discussion. It is found that the low-energy states are mainly from the orbitals of Ag and Bi atoms, which lie in the (110) mirror plane. Notably, the structure does not preserve an inversion center. Besides the (110) mirror plane $M_x$ (here we take [110] as $x$-direction), two other important crystalline symmetries are the three-fold rotation $C_{3z}$ and the two-fold screw rotation $S_{2z}=\{C_{2z}|00\frac{1}{2}\}$.

We performed first-principles calculations based on the density functional theory (DFT). The calculation details are presented in the Supplemental Material~\cite{SI}. Figure~\ref{fig3}(a) shows the plot of calculated band structure of CaAgBi with spin-orbit coupling (SOC) included. One observes that the material is a semimetal, and the bands around Fermi level are dominated by the Ag-$5s$ and Bi-$6p$ orbitals. Normally, Ag-$5s$ orbitals have a higher energy than Bi-$6p$ orbitals. However, around $\Gamma$-point, one observes a band inversion feature with Ag-$5s$ lower than Bi-$6p$. Importantly, along the $k_z$-axis ($\Gamma$-A), the three low-energy bands cross pairwisely and linearly at three discrete points, which we label as $D_1$, $D_2$, and $D_3$, as shown in Fig.~\ref{fig3}(b). $D_1$ and $D_2$ are on the rotational axis, which are accidental crossing points originated from the band inversion at $\Gamma$. In contrast, $D_3$ is pinned at A-point on the BZ boundary, which is an essential crossing point dictated by the space group symmetry. One notes that each of the three crossing bands are two-fold degenerate (with SOC) on $\Gamma$-A, which is caused by the non-commutativity between $S_{2z}$ (or $C_{3z}$) and $M_x$ symmetries along this path. Thus each crossing point $D_i$ ($i=1,2,3$) is of four-fold degeneracy, corresponding to a Dirac point.

Let's first consider the two accidental Dirac points $D_1$ and $D_2$. On $\Gamma$-A path, the point group symmetry is of $C_{6v}$, and the three doubly-degenerate crossing bands are found to form three \emph{distinct} two-dimensional irreducible representations $E_{1/2}$, $E_{3/2}$ and $E_{5/2}$ of the $C_{6v}$ double group (see Fig.~\ref{fig3}(b)). Therefore, these bands cannot hybridize along this line, and their crossing points are protected and of linear type. The symmetry protection is lost for $k$-points deviating from the rotational axis, hence each crossing should be an isolated point rather than a line. Another important observation is that while the slopes of the two crossing bands at $D_1$ have opposite signs, the two bands at $D_2$ have the same sign. Thus $D_1$ and $D_2$ are type-I and type-II Dirac points, respectively.

The dispersions around $D_1$ and $D_2$ in the $k_x$-$k_y$ plane also show interesting features. As shown in Fig.~\ref{fig3}(c,d), the bands completely split in directions different from the $k_z$-axis, which can be understood by noticing that the two-fold band degeneracy on $\Gamma$-A cannot be maintained at generic $k$-points in the $k_x$-$k_y$ plane because such points are not invariant under the rotational symmetries $S_{2z}$ and $C_{3z}$. This is in contrast to most proposed Dirac semimetals, where the two-fold band degeneracy is maintained at any $k$-point due to the presence of inversion symmetry (in combination with time reversal symmetry).

\begin{figure}[t!]
\includegraphics[width=9cm]{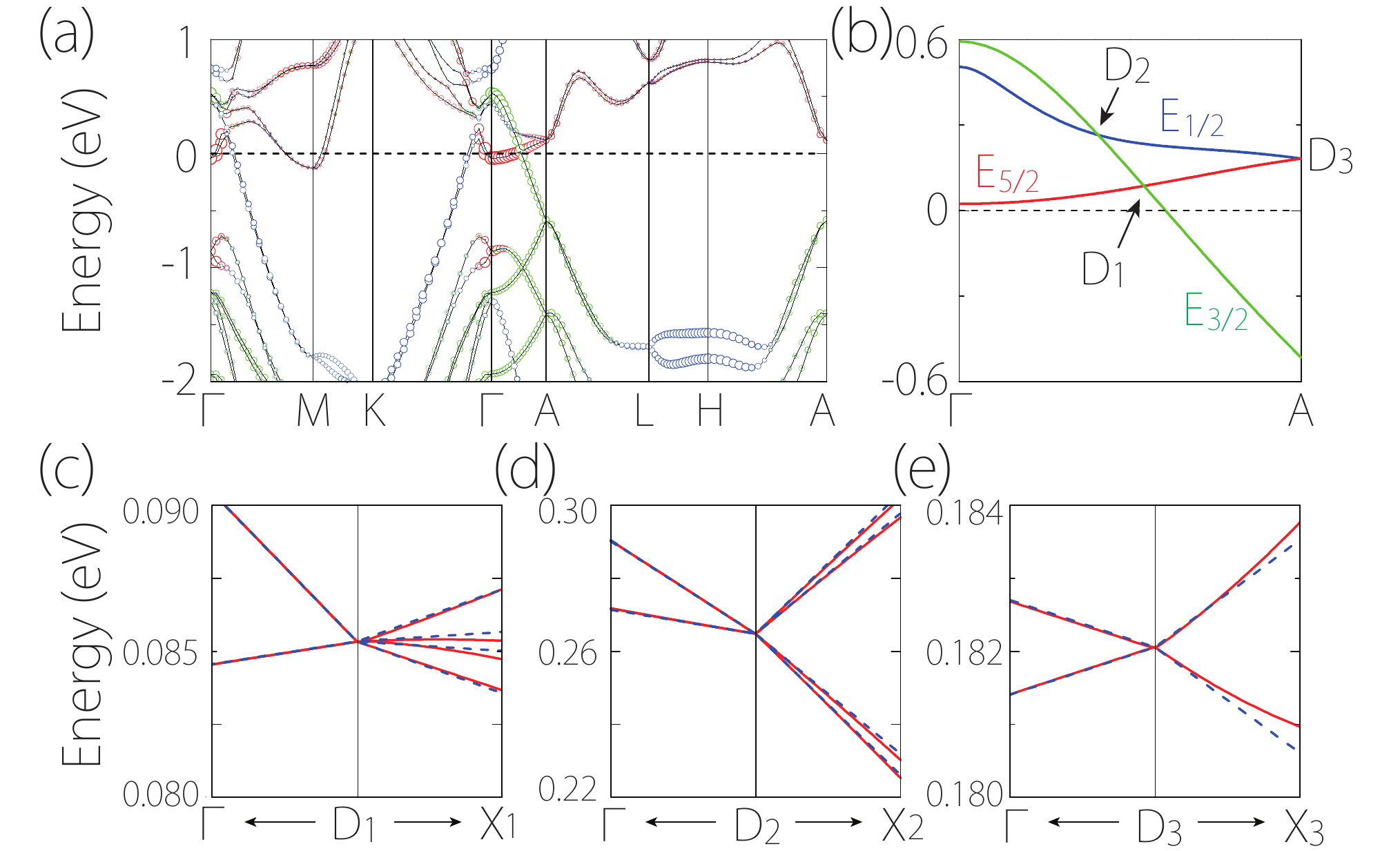}
\caption{(a) Band structure of CaAgBi (SOC included). The colored circles indicate the weight of orbital projection onto Ag-$s$ (red), Bi-$p_{x/y}$ (green), and Bi-$p_z$ (blue) orbitals. (b) Enlarged low-energy band structure along $\Gamma$-A, showing three Dirac points. (c-e) Dispersions around the three Dirac points by first-principles calculations (red solid lines) and $k\cdot p$ model fitting (blue dashed lines). Here $D_i$-$X_i$ ($i=1,2,3$) is along the $k_x$-direction perpendicular to $k_z$-axis.  }
\label{fig3}
\end{figure}

To characterize the low-energy Dirac fermions, we construct the $k\cdot p$ effective models around each Dirac point, subjected to the symmetry constraints~\cite{Bradley1972}. For point $D_1$, it is at the intersection between $E_{5/2}$ band and $E_{3/2}$ band on $\Gamma$-A. Using the four-states with such symmetries at $D_1$ as basis, the Hamiltonian around $D_1$ up to linear order in $\bm q$ takes the form
\begin{equation}\label{D1}
\mathcal{H}_{D_1}=C_1q_z+\left[
          \begin{array}{cccc}
            C_2 q_z & B^* q_+ & iAq_- & 0 \\
            B q_- & -C_2q_z & 0 & 0 \\
            -iAq_+ & 0 & C_2q_z & -B^* q_- \\
            0 & 0 & -B q_+ & -C_2q_z \\
          \end{array}
        \right],
\end{equation}
where the wave-vector $\bm q$ and the energy are measured from $D_1$, $q_\pm=q_x\pm iq_y$, the model parameters $A$, $C_1$, $C_2$ are real, and $B$ is complex. The first term in Eq.~(\ref{D1}) represents a tilt of spectrum. Figure~\ref{fig3}(c) shows the fitting of this model to the DFT band structure. Particularly, we find that $|C_2|>|C_1|$, such that the point is of type-I with dispersion $\varepsilon=(C_1\pm C_2)q_z$ along $k_z$-direction. Similarly, we can find the effective model around point $D_2$, which is at the intersection between $E_{1/2}$ and $E_{3/2}$ bands. We find that the obtained Hamiltonian $\mathcal{H}_{D_2}$ has the same form as $\mathcal{H}_{D_1}$, and only the values of the parameters are different. Most importantly, now $|C_2|<|C_1|$. The dispersion along $k_z$-direction is dominated by the tilt term, so that $D_2$ is a type-II Dirac point (see Fig.~\ref{fig3}(d)).

Next, we come to the essential Dirac point at A. Its essential character can be argued as follows. The bands along $\Gamma$-A can be chosen as $S_{2z}$ eigenstates. Since $(S_{2z})^2=-e^{-ik_z}$ (the minus sign is from $2\pi$ spin rotation, and $k_z$ is measured in unit of $1/c$), the $S_{2z}$ eigenvalues are given by
\begin{equation}
s =\pm i e^{-ik_z/2}.
\end{equation}
Meanwhile, we have $\{S_{2z}, M_x\}=0$, so the two states $|s\rangle$ and $M_x|s\rangle$ form a degenerate pair with \emph{opposite} $S_{2z}$ eigenvalues, where $|s\rangle$ denotes an eigenstate of $S_{2z}$ with eigenvalue $s$. This explicitly demonstrates  the two-fold band degeneracy along $\Gamma$-A, as we claimed before. Note that A is invariant under time reversal operation ($\mathcal{T}$), such that at A, each state $|s\rangle$ has a Kramers degenerate partner $\mathcal{T}|s\rangle$. Since $s=\pm 1$ at point A ($k_z=\pi$), $\mathcal{T}|s\rangle$ shares the same $S_{2z}$ eigenvalue as $|s\rangle$. Consequently, the four states $\{|s\rangle, M_x|s\rangle, \mathcal{T}|s\rangle, \mathcal{T}M_x|s\rangle\}$  at A are linearly independent and must be degenerate with the same energy. Thus the nonsymmorphic space group symmetry necessitates the four-fold degeneracy at A-point. The effective model up to linear order in $\bm q$ expanded around $D_3$ is obtained as
\begin{equation}\label{D3}
\mathcal{H}_{D_3}=-v_\bot \sigma_y\tau_z q_x+v_\bot \sigma_x\tau_z q_y+v_z\sigma_z\tau_x q_z.
\end{equation}
Here $\bm q$ and energy are measured from $D_3$, $\sigma_i$ and $\tau_i$ are the Pauli matrices, and model parameters $v_\bot$ and $v_z$ correspond to the Fermi velocities. Figure~\ref{fig3}(e) shows the fitting of DFT band structure using model (\ref{D3}). Compared with model (\ref{D1}) for the two accidental Dirac points, model (\ref{D3}) has less free parameters due to the additional time reversal symmetry at A. Consequently, the dispersion around $D_3$ is different from the other two in that: (i) there is no energy tilt term (hence it must be type-I); and (ii) the bands in $k_z=\pi$ plane are still two-fold degenerate without splitting.

\begin{figure}[t!]
\includegraphics[width=9cm]{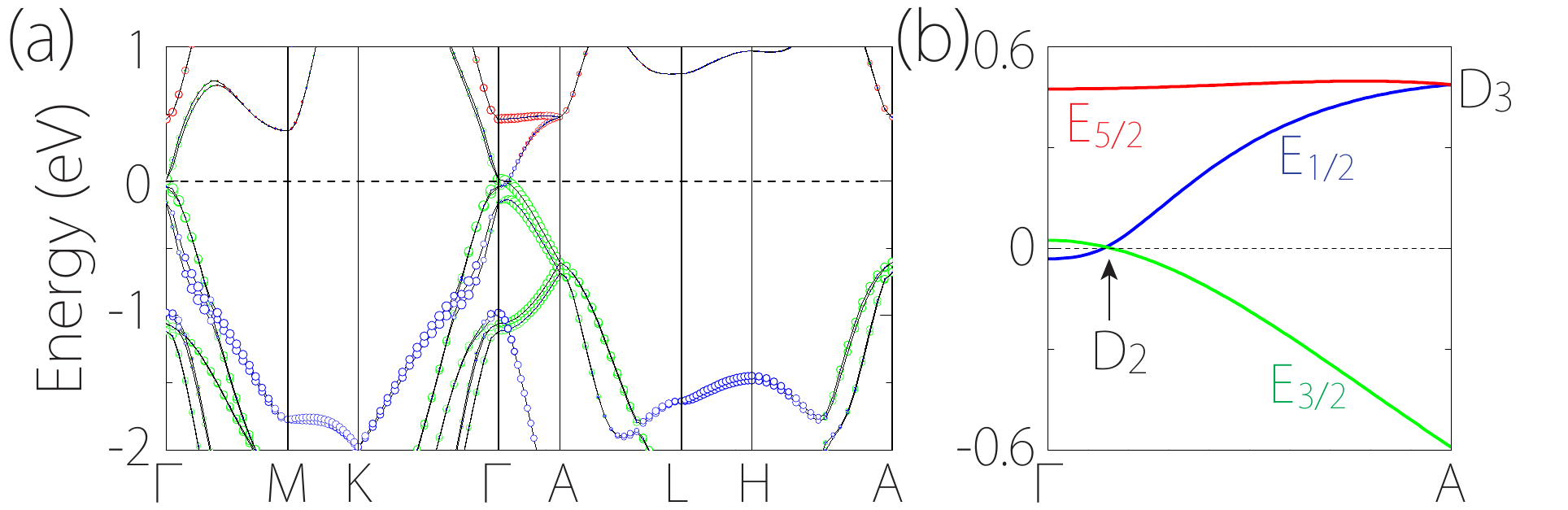}
\caption{(a) Band structure of CaAgAs (SOC included). The colored circles indicate the weight of orbital projection onto Ag-$s$ (red), As-$p_{x/y}$ (green), and As-$p_z$ (blue) orbitals. (b) Enlarged low-energy band structure along $\Gamma$-A.  }
\label{fig4}
\end{figure}

We further clarify the distinction between the accidental Dirac points ($D_1$, $D_2$) and the essential Dirac point ($D_3$). The realization of accidental Dirac points requires band inversion. In the current system, this can be characterized by two $\mathbb{Z}_2$  indices $\zeta_1$ and $\zeta_2$. Here $\zeta_1=\text{sgn}(\Delta_{\Gamma,1}\Delta_{\text A, 1})\in \{+1,-1\}$, with $\Delta_{k,1}=\varepsilon_{5/2}(k)-\varepsilon_{3/2}(k)$, where $\varepsilon_i(k)$ ($i=5/2$ or $3/2$, and $k=\Gamma$, A) denotes the energy of the $E_i$ band at  point $k$. Evidently, $\zeta_1=-1$ corresponds to the nontrivial case with band inversion, leading to the $D_1$ Dirac point. Similarly,
$\zeta_2$ is defined with $E_{5/2}$ replaced by $E_{1/2}$, and characterizes the band inversion between $E_{1/2}$ and $E_{3/2}$ bands for the formation of $D_2$. The nontriviality  of $\zeta_i$ ($i=1,2$) is accidental, depending on the material. For example, in CaAgAs, which  is isostructural to CaAgBi, we find that $\zeta_1=+1$ (see Fig.~\ref{fig4}), indicating that the band inversion between $E_{3/2}$ and $E_{5/2}$ is removed. Therefore, $D_1$ point is eliminated. Meanwhile, $\zeta_2$ remains $-1$, so $D_2$ is kept, but it may ultimately be eliminated by a symmetry-preserving strain. In contrast, $D_3$ is invariably present in all these cases, since it is only determined by symmetry.

\begin{figure}[h!]
\includegraphics[width=7.5cm]{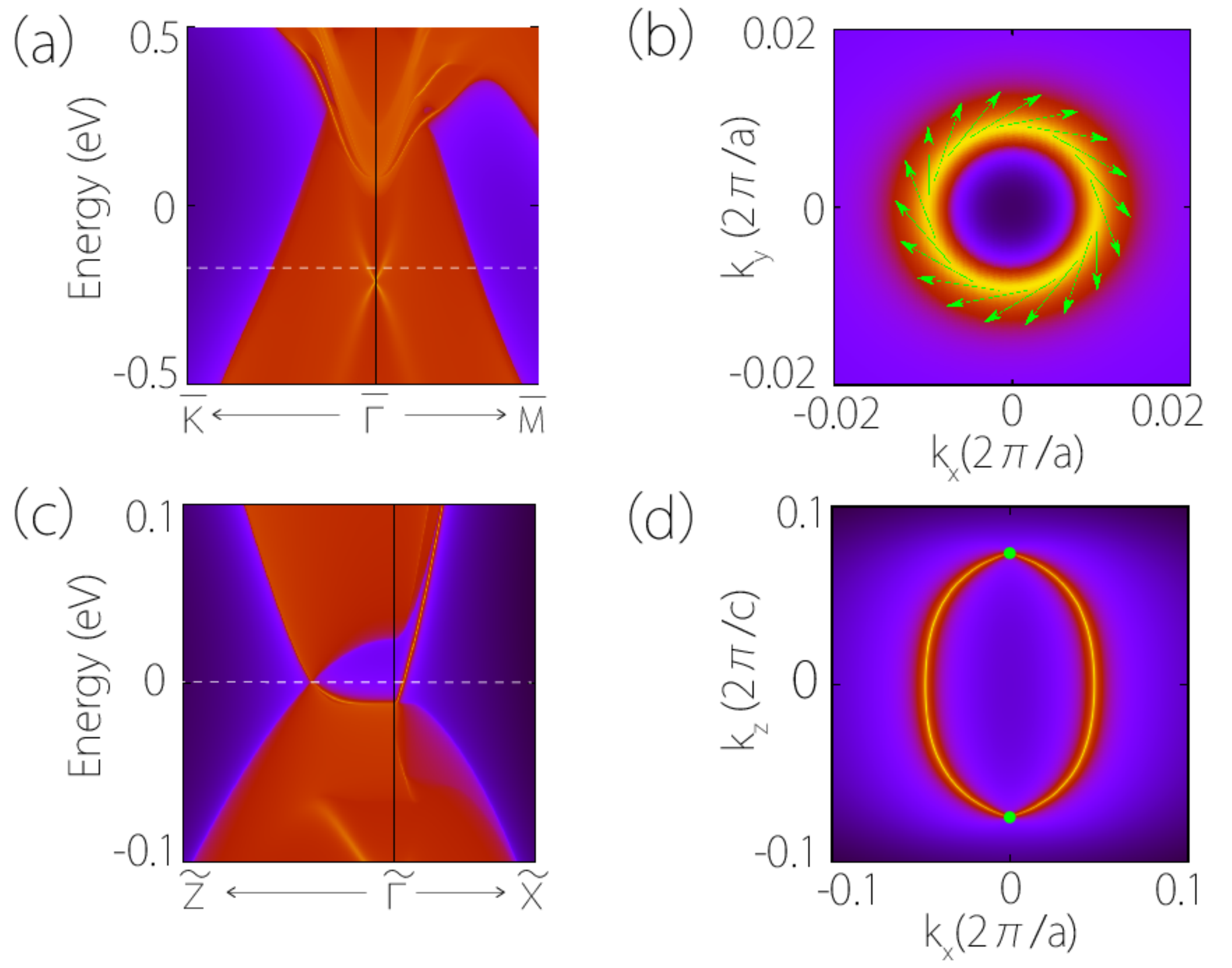}
\caption{(a) Projected spectrum on (001) surface for CaAgBi, showing Dirac-cone surface states buried in the bulk valence band. (b) Surface states at -0.18 eV (marked in (a)), showing a left-handed spin-momentum locking. (c) Projected spectrum on (010) surface for CaAgAs, showing (d) surface Fermi arcs connecting a pair of projected Dirac points (green dots). }
\label{fig5}
\end{figure}

Band topology and intrinsic anisotropy lead to interesting surface spectra of CaAgBi-family materials, with distinct topological surface states on different crystal surfaces. Here the Dirac points are all residing on $k_z$-axis. They are projected onto the same image point on (001) surface. Since the band is inverted at $\Gamma$-point in the bulk, we expect to have Dirac-cone type surface states similar to the topological insulator case. In Fig.~\ref{fig5}(a), we indeed find such surface states buried in the bulk valence bands. Figure~\ref{fig5}(b) shows the constant energy slice above the surface Dirac point, at which the surface states exhibit a left-handed spin helicity, similar to the pattern for strong topological insulators~\cite{Sheng_TlN,2017arXiv170309040S}. Meanwhile, the Dirac points get projected to different image points on the side surfaces. For (010) surface, we find Fermi arcs connecting a pair of projected Dirac points, as shown in Fig.~\ref{fig5}(c,d) for CaAgAs (for CaAgBi, the arcs are buried in the bulk bands~\cite{SI}). These Fermi arcs are in fact protected by a bulk $\mathbb{Z}_2$ invariant. Consider the $\mathcal{T}$-invariant planes $k_z=0$ and $k_z=\pi$. The two planes are fully gapped hence each has a well-defined 2D $\mathbb{Z}_2$ invariant. Using the Wilson loop method~\cite{YuR_Wilson,PhysRevB.83.035108,Wu2017}, we find that $\mathbb{Z}_2=1$ for $k_z=0$ plane, whereas $\mathbb{Z}_2=0$ for $k_z=\pi$ plane. Thus a Kramers pair of surface states must exist on the $k_z=0$ path for the side surfaces, ensuring the presence of a pair of surface Fermi arcs.

\begin{figure}[t!]
\includegraphics[width=8.5cm]{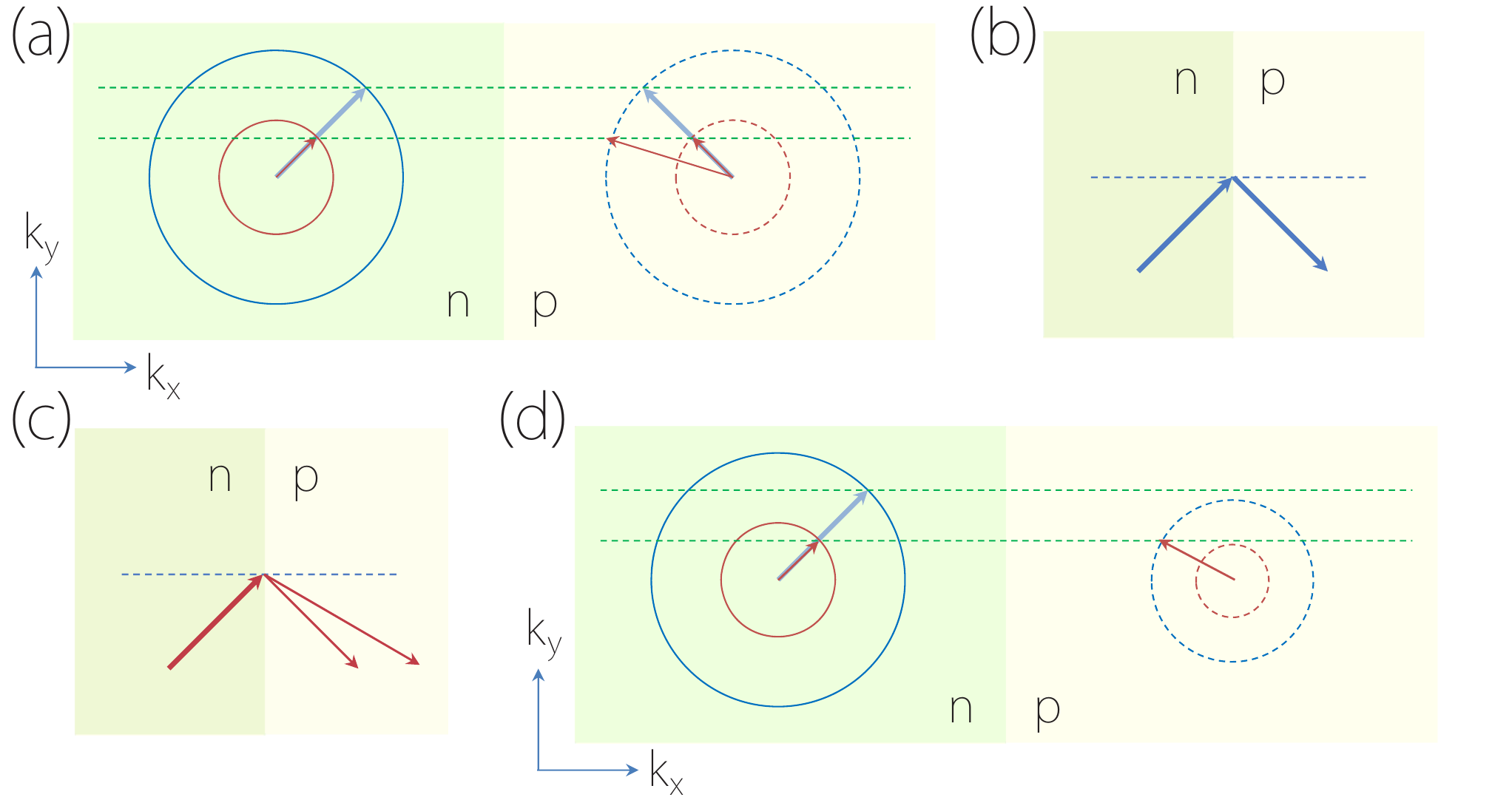}
\caption{Schematic figures showing Dirac fermion transport at a p-n junction along $x$. (a) Fermi circles (near $D_1$ point) for two sides of the junction. (b) Transmitted carriers show negative refraction at the interface. (c) Double refraction occurs for an incident state with two transmitted states (marked in (a)). (d) At certain range of bias voltage and incident angle, only states from the inner Fermi circle get transmitted.     }
\label{fig6}
\end{figure}

As mentioned, there exist a large family of ternary compounds with stuffed Wurtzite structure. Many members of this family share the same band features as discussed above (see Supplemental Material~\cite{SI}), exhibiting a hybrid Dirac semimetal phase with multiple Dirac points. Since the different kind of Dirac points are at different energies, it is possible to control the type of Dirac fermions in a single material by tuning the doping level, probing each kind separately. Alternatively, one may choose the proper material from the family which has the desired Dirac point most close to the Fermi level.

A salient feature of these materials is the broken inversion symmetry, which makes the Dirac bands completely split except along the $k_z$-axis. This could lead to a new double negative refraction effect at a p-n junction of such material. Consider the case when the Fermi level is most close to the $D_1$ point. Consider ballistic transport across a p-n junction along $x$-direction, and we focus on the carrier dynamics in the $x$-$y$ plane (the transverse momenta $k_y$ and $k_z$ are conserved). As illustrated in Fig.~\ref{fig6}(a), with a constant $k_z$, there are two Fermi circles around the Dirac point, due to the band splitting. Carriers on each Fermi circle is chiral and undergo refractions at the junction interface, similar to that in graphene~\cite{Cheianov:2007in}. For p-n junction, this leads to negative refraction effect (Fig.~\ref{fig6}(b)), and for incident state with two transmitted states, there are two negative refraction paths (see Fig.~\ref{fig6}(c)). In addition, by controlling the bias voltage and the incident angle of an incident electron beam, one can have transmission of carriers from only one of the Fermi circle (see Fig.~\ref{fig6}(d)), whereas carriers from the other circle are totally reflected. This could be useful in electron optics to spatially separate the chiral carriers.

In conclusion, we have demonstrated a novel hybrid Dirac semimetal phase in CaAgBi-family materials. The low-energy band structure features three kinds of Dirac points: two pairs of accidental Dirac points with type-I and type-II dispersions, and a single essential Dirac point dictated by nonsymmorphic symmetry. We clarify their symmetry protections and characterize their individual low-energy Dirac fermions.
Distinct from centrosymmetric Dirac semimetals, the Dirac bands here completely split  due to the lack of inversion center. We show that this feature leads to interesting transport features at a p-n junction. We also find that different crystalline surfaces possess different topological surface states. The bulk Dirac points and the surface states can be experimentally probed by the angle-resolved photoemission spectroscopy (ARPES).
Our findings make it possible to study multiple types of Dirac fermions and their interplays within a single material system.

%\begin{acknowledgements}
%The authors thank D.L. Deng for valuable discussion. This work is supported by the NSF of China (No. 11504013), Singapore MOE Academic Research Fund Tier 2 (MOE2015-T2-2-144) and Tier 1 (SUTD-T1-2015004).
%\end{acknowledgements}

%\begin{references}
%\end{references}

\bibliography{CaAgBi_ref}

%\newpage

\end{document}